\documentclass[
aps,
prd,
10pt,
showpacs,
amsmath,
amssymb,
twocolumn,
nofootinbib,
nobibnotes,
preprintnumbers
]{revtex4-1}
\usepackage{float}
\usepackage{mathrsfs,amsfonts}
\usepackage{mathtools}
\usepackage{tensor}
\usepackage{nicefrac}
\usepackage{url}
\usepackage[hyperindex,breaklinks]{hyperref}
\pdfoutput=1

\begin{document}
\allowdisplaybreaks[1]
\title{Quantum equivalence of $f(R)$ gravity and scalar-tensor theories}
\author{Michael S. Ruf}
\author{Christian F. Steinwachs}
\email{christian.steinwachs@physik.uni-freiburg.de}
\affiliation{Physikalisches Institut, Albert-Ludwigs-Universit\"at Freiburg,\\
Hermann-Herder-Stra\ss e~3, 79104 Freiburg, Germany}
%
%

\begin{abstract}
We investigate whether the classical equivalence of $f(R)$ gravity and its formulation as scalar-tensor theory still holds at the quantum level.
We explicitly compare the corresponding one-loop divergences and find that the equivalence is broken by off-shell quantum corrections, but recovered on-shell.
\end{abstract}

%
%
\maketitle									  
%
%


\section{Introduction}
Scalar-tensor-theories and \mbox{$f(R)$ theories} have important applications in cosmological models, which describe the early and late time acceleration of the universe \cite{Sotiriou:2008rp,DeFelice:2010aj,Nojiri:2010wj,Clifton:2011jh,Nojiri:2017ncd}. 
Conceptually, scalar-tensor theories and \mbox{$f(R)$ theories} are different.
While scalar-tensor theories introduce scalar ``matter'' degrees of freedom to the unmodified Einstein-Hilbert action, \mbox{$f(R)$ theories} correspond to a modification of the underlying gravitational theory without adding any new matter degrees of freedom.

In contrast to Einstein's theory, which involves at most second derivatives of the metric field, a generic \mbox{$f(R)$ theory} is a fourth order theory. 
Beside the massless spin two graviton, present in the spectrum of Einstein's theory, higher derivatives propagate additional degrees of freedom \cite{Stelle:1976gc,Stelle:1977ry}.
Generically, fourth order theories of gravity lead to an additional massive spin zero degree of freedom, the ``scalaron'' and an additional massive spin two ghost \cite{Stelle:1976gc, Stelle:1977ry, Starobinsky:1980te}.  
Among higher derivative theories of gravity, \mbox{$f(R)$ gravity} is special.
Despite being a fourth order theory, \mbox{$f(R)$ gravity} does not propagate the ghost and therefore avoids the classical Ostrogradski instability and the associated problems with unitarity violation at the quantum level \cite{Stelle:1976gc,Stelle:1977ry,Woodard:2006nt}.

Beside the aforementioned differences between the interpretation of scalar-tensor theories and \mbox{$f(R)$ theories}, both introduce an additional scalar degree of freedom and share many similarities.
For example, the predictions of two natural and successful models of inflation, Starobinsky's $R^2$-model \cite{Starobinsky:1980te} and non-minimal Higgs inflation \cite{Bezrukov:2007ep,Barvinsky:2008ia,Bezrukov:2008ej,DeSimone:2008ei,Barvinsky:2009fy,Bezrukov:2009db,Barvinsky:2009ii, Bezrukov:2010jz}, are almost indistinguishable for strong non-minimal coupling \cite{Barvinsky:2008ia, Bezrukov:2011gp, Kehagias:2013mya}.
This is a manifestation of the fact that \mbox{$f(R)$ gravity} admits a classically equivalent formulation as a scalar-tensor theory.\footnote{In contrast, not all scalar-tensor theories can be reformulated as $f(R)$ theory.}

In this paper we investigate whether this classical equivalence between \mbox{$f(R)$ gravity} and scalar-tensor still holds at the quantum level. 
The one-loop divergences $\varGamma_{1}^{f}$ for \mbox{$f(R)$ gravity} have been calculated recently on an arbitrary background \cite{wefR}.
Likewise, the one-loop divergences $\hat{\varGamma}_{1}^{\text{\tiny EF}}$ for a scalar field minimally coupled to gravity have been calculated in \cite{Barvinsky:1993zg, Kamenshchik:2014waa}.\footnote{A ``hat'' indicates that the corresponding quantity is expressed in terms of the Einstein frame fields $(\hat{g}_{\mu\nu},\hat{\varphi})$.}

\begin{figure}[H]
      \centering
      \includegraphics[]{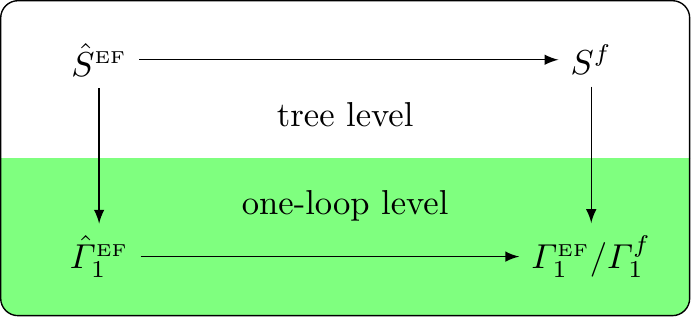}
      \caption{Transition between different formulations}
      \label{Fig1}
\end{figure}

We use the transformation between the classical action of a scalar-tensor theory in the Einstein frame parametrization $S^{\text{\tiny EF}}$  and its \mbox{$f(R)$ formulation} $S^{f}$ to transform $\hat{\varGamma}_{1}^{\text{\tiny EF}}$ to its \mbox{$f(R)$ formulation} $\varGamma_{1}^{\text{\tiny EF}}$. We then compare $\varGamma_{1}^{\text{\tiny EF}}$ to the one-loop result $\varGamma_{1}^{f}$, obtained directly in the \mbox{$f(R)$ formulation}.
The question of quantum equivalence can be summarized pictorially by the question of whether the diagram in \mbox{FIG. \ref{Fig1}} commutes or not.
The question of the equivalence between \mbox{$f(R)$ gravity} and its scalar-tensor formulation is related to the similar question of equivalence between different field parametrizations in scalar-tensor theories.
In particular, there is a rather old but still ongoing debate about the equivalence of the so-called Jordan frame and Einstein frame parametrizations used in cosmological models \cite{Dicke:1961gz,Bergmann:1968ve,Magnano:1993bd,Faraoni:1999hp,Nojiri:2000ja,Capozziello:2010sc,Calmet:2012eq,Steinwachs:2013tr,Prokopec:2013zya,Chiba:2013mha,Kamenshchik:2014waa,Postma:2014vaa,Jarv:2014hma,Domenech:2015qoa, Herrero-Valea:2016jzz,Kamenshchik:2016gcy,Pandey:2016jmv,Jarv:2016sow,Bahamonde:2017kbs,Bhattacharya:2017pqc,Karam:2017zno}.
The quantum equivalence between the Jordan frame and the Einstein frame has been investigated in \cite{Kamenshchik:2014waa}, by an explicit comparison of the one-loop divergences -- similar to the analysis in this paper. 
Beside the similarity to \cite{Kamenshchik:2014waa} in the method of comparison, the underlying problem for $f(R)$ gravity is different.

The transition between Einstein frame and Jordan frame maps a second order scalar-tensor theory of the two fields $(\hat{g}_{\mu\nu},\hat{\varphi})$ to a second order scalar-tensor theory of the two fields $(g_{\mu\nu},\varphi)$.
In contrast, the transition between the Einstein frame scalar-tensor theory and \mbox{$f(R)$ gravity} maps a second order theory of the fields $(\hat{g}_{\mu\nu},\hat{\varphi})$ to a purely gravitational fourth order theory of one field $g_{\mu\nu}$.
Therefore, the explicit transformation rules are not only non-linear but also involve derivatives.

The paper is structured as follows.
In Sec.~\ref{Sec:ScalarTensor}, we present the Jordan frame and the Einstein frame formulation of scalar-tensor theories and provide the result for the one-loop divergences of the latter.
In Sec.~\ref{Sec:fRGravity}, we discuss \mbox{$f(R)$ gravity} and its one-loop divergences.
In Sec.~\ref{Sec:Transformations}, we derive the explicit transformation laws for the transition from the Einstein frame scalar-tensor formulation to \mbox{$f(R)$ gravity}.
In Sec.~\ref{Sec:Comparison}, we transform the one-loop divergences for the Einstein frame scalar-tensor formulation to its $f(R)$ formulation and compare the result to the one-loop divergences obtained directly for \mbox{$f(R)$ gravity}.
In Sec.~\ref{Sec:Conclusion}, we summarize our main results and discuss their implications.


\section{Scalar tensor theory}\label{Sec:ScalarTensor}

The Euclidean action of a scalar-tensor theory for a single scalar field $\varphi$ can be parametrized by three arbitrary functions $U(\varphi)$, $G(\varphi)$ and $V(\varphi)$, 
\begin{align}
S^{\text{\tiny JF}}[g,\varphi]=\int{\rm d}^4x\,g^{\nicefrac{1}{2}}\left(-UR+\frac{G}{2}g^{\mu\nu}\partial_{\mu}\varphi\,\partial_{\nu}\varphi+V\right)\,.\label{ActJF}
\end{align}
This representation of scalar-tensor theories is called Jordan frame (JF) parametrization.
Performing a conformal transformation of the metric field $g_{\mu\nu}$ and a redefinition of the scalar field $\varphi$, 
\begin{align}
\tensor{\hat{g}}{_\mu_\nu}=\frac{U}{U_0}\tensor{g}{_\mu_\nu}\,,\;\; \left(\frac{\partial\hat{\varphi}}{\partial\varphi}\right)^2=\left(\frac{U_0}{U}\right)\frac{G\,U+3\,(U_{1})^2}{U}\,,\label{EFTrafo}
\end{align}
where $U_1=\partial U(\varphi)/\partial\varphi$, the action \eqref{ActJF} transforms into 
\begin{align}
\hat{S}^{\text{\tiny EF}}[\hat{g},\hat{\varphi}]=\int{\rm d}^4x\,\hat{g}^{\nicefrac{1}{2}}\left(-U_0\,\hat{R}+\frac{1}{2}\hat{g}^{\mu\nu}\partial_{\mu}\hat{\varphi}\,\partial_{\nu}\hat{\varphi}+\hat{V}\right)\,.\label{ActEF}
\end{align}
The action \eqref{ActEF} resembles the Einstein-Hilbert action for $\hat{g}_{\mu\nu}$ with a minimally coupled scalar field $\hat{\varphi}$. Consequently, the parametrization in terms of the variables $(\hat{g}_{\mu\nu},\hat{\varphi})$ is called Einstein frame (EF).
Here, $U_0$  is a constant, usually identified with the Planck mass ${U_0=M_{\mathrm{P}}^2/2}$ and $\hat{V}$ is the EF potential, defined by
\begin{align}
\hat{V}(\hat{\varphi})\coloneqq U_0^2\,\frac{V(\varphi)}{U^2(\varphi)}\Big|_{\varphi=\hat{\varphi}}\,.\label{EFPot}
\end{align}
Extremizing the EF action \eqref{ActEF} with respect to $\hat{g}_{\mu\nu}$ and $\hat{\varphi}$ gives rise to the Einstein equation for $\hat{g}_{\mu\nu}$ and the Klein-Gordon equation for $\hat{\varphi}$,
\begin{align}
\tensor{\hat{R}}{_\mu_\nu}-\frac{1}{2}\hat{R}\,\tensor{\hat{g}}{_\mu_\nu}&=\frac{1}{2U_0}\tensor*{T}{^{\hat\varphi}_\mu_\nu}\,,
\qquad\hat{\Delta}\,\hat{\varphi}=-\hat{V}_1\,.\label{EinsteinKleinGordonEq}
\end{align}
Here, $\hat{\Delta}\coloneqq-\tensor{\hat{g}}{^\mu^\nu}\tensor{\hat{\nabla}}{_{\mu}}\!\tensor{\hat{\nabla}}{_{\nu}}$ is the Laplacian and $\tensor*{T}{^{\hat\varphi}_\mu_\nu}$ is the scalar field energy momentum tensor
\begin{align}
\tensor*{T}{^{\hat\varphi}_\mu_\nu}\coloneqq{}&\tensor{\partial}{_\mu}\hat{\varphi}\,\tensor{\partial}{_\nu}\hat{\varphi}-\frac{1}{2}\tensor{\hat{g}}{_\mu_\nu}\left(\tensor{\hat{g}}{^\alpha^\beta}\tensor{\partial}{_\alpha}\hat{\varphi}\,\tensor{\partial}{_\beta}\hat{\varphi}+2\hat{V}\right)\,.
\end{align}
We denote derivatives of the EF potential $\hat{V}$ with respect to the EF scalar field $\hat{\varphi}$ by
\begin{align}
\hat{V}_n\coloneqq\frac{\partial^n \hat{V}}{\partial\hat{\varphi}^n}\,.
\end{align}
The calculation of the one-loop effective action requires a proper gauge fixing. In \cite{Barvinsky:1993zg,Kamenshchik:2014waa}, the background covariant de Donder gauge condition is used
\begin{align}
\tensor{\hat\chi}{^\alpha}[\hat{g},\hat{h}]&=-\tensor{\hat{g}}{^\alpha^\mu}\tensor{\hat{g}}{^\beta^\nu}\left(\tensor{\hat\nabla}{_\beta}\tensor{\hat{h}}{_\mu_\nu}-\frac{1}{2}\tensor{\hat\nabla}{_\mu}\tensor{\hat{h}}{_\beta_\nu}\right)\label{DeDonder}\,.
\end{align}
The covariant derivative $\hat{\nabla}_{\mu}$ is defined with respect to the metric $\hat{g}_{\mu\nu}$.
The one-loop divergences for the EF action \eqref{ActEF}, obtained in \cite{Barvinsky:1993zg, Kamenshchik:2014waa}, read\footnote{The same result can be obtained from the one-loop divergences for the JF action \eqref{ActJF}, calculated in \cite{Shapiro:1995yc, Steinwachs:2011zs}, in the limit $U=U_0$, $G=1$, by setting $V=\hat{V}$, $g_{\mu\nu}=\hat{g}_{\mu\nu}$ and $\varphi=\hat{\varphi}$. The model \eqref{ActJF} with $G=1$ has also been considered within the exact functional renormalization group in \cite{Narain:2009fy, Percacci:2015wwa}. Note that the results in \cite{Barvinsky:1993zg, Kamenshchik:2014waa} are obtained in Lorentzian signature. Their transformation to the Euclidean version \eqref{OneLoopGammaDivEF} involves a global minus sign.}
\begin{widetext}
\begin{align}
\hat{\varGamma}_{1}^{\text{\tiny EF}}\big|^{\mathrm{div}}=\frac{1}{32\,\pi^2\varepsilon}\int{\rm d}^4x\,\hat{g}^{\nicefrac{1}{2}}&\Bigg\{-\frac{71}{60}\hat{\mathcal{G}}-\frac{43}{60}\hat{R}_{\mu\nu}\hat{R}^{\mu\nu}-\frac{1}{40}\hat{R}^2+\frac{1}{6}\hat{R}\hat{V}_{2}-\frac{1}{2}\left(\hat{V}_{2}\right)^2\nonumber\\
&+U_0^{-1}\left[\frac{13}{3}\hat{R}\,\hat{V}+\frac{1}{3}\hat{R}\,\left(\partial_\mu\hat{\varphi}\partial^{\mu}\hat{\varphi}\right)+2\,\left(\hat{V}_{1}\right)^2+2\,\hat{V}_{2}\left(\partial_\mu\hat{\varphi}\partial^{\mu}\hat{\varphi}\right)\right]\nonumber\\
&-U_0^{-2}\left[5\,\hat{V}^2+\hat{V}\left(\partial_\mu\hat{\varphi}\partial^{\mu}\hat{\varphi}\right)+\frac{5}{4}\left(\partial_\mu\hat{\varphi}\partial^{\mu}\hat{\varphi}\right)^2\right]\Bigg\}\,.\label{OneLoopGammaDivEF}
\end{align}
\end{widetext}
The Gauss-Bonnet term in the EF parametrization is defined as
\begin{align}
\mathcal{\hat{G}}\coloneqq{}&\tensor{\hat{R}}{_{\mu\nu\rho\sigma}}\tensor{\hat{R}}{^{\mu\nu\rho\sigma}}-4\tensor*{\hat{R}}{_{\mu\nu}}\tensor*{\hat{R}}{^{\mu\nu}}+\hat{R}^2\,.\label{GaussBonnetEF}
\end{align}
It is understood that the indices in \eqref{OneLoopGammaDivEF} and \eqref{GaussBonnetEF} are raised and lowered with the metric $\hat{g}_{\mu\nu}$.


\section{$f(R)$ gravity}\label{Sec:fRGravity}
The Euclidean action functional for \mbox{$f(R)$ theories} is given by
\begin{align}
S^{f}[g]=-\int\mathop{}\!\mathrm{d}^{4}x\,g^{\nicefrac{1}{2}}\,f(R).\label{fRAct}
\end{align}
We denote derivatives of the function $f$ with respect to its argument by a subindex 
\begin{align}
f_{n}\coloneqq\frac{\partial^{n}\!f(R)}{\partial R^n}\,.
\end{align}
The extremal is defined as
\begin{align}
\tensor{\mathscr{E}}{_\mu_\nu}\coloneqq{}&  \tensor{g}{_\mu_\alpha}\tensor{g}{_\nu_\beta}\,g^{-\nicefrac{1}{2}}\,\frac{\delta S[g]}{\delta \tensor{g}{_\alpha_\beta}}\nonumber\\
={}&-\Delta f_1\,\tensor{g}{_\mu_\nu}-\nabla_{\mu}\nabla_{\nu}f_1+f_1\tensor*{R}{_\mu_\nu}-\frac{1}{2}f\tensor{g}{_\mu_\nu}\,.\label{EOM}
\end{align}
The classical equations of motion are satisfied, if $\tensor{\mathscr{E}}{_\mu_\nu}=0$.
The trace of the extremal reads
\begin{align}
\mathscr{E}\coloneqq \tensor{g}{^\mu^\nu}\mathscr{E}_{\mu\nu}=-3\,\Delta f_1+Rf_1-2f\,.\label{OnShellScalar}
\end{align}
We also define the rescaled extremal and its trace
\begin{align}
\tensor{E}{_\mu_\nu}\coloneqq\frac{\tensor{\mathscr{E}}{_\mu_\nu}}{(-f_1)},\qquad E\coloneqq\tensor{g}{^\mu^\nu}E_{\mu\nu}=\frac{\mathscr{E}}{(-f_1)}\,.
\end{align}
Both, $E_{\mu\nu}$ and $E$, are homogeneous functions of degree zero in $f$ and its derivatives $f_n$.
The one-loop divergences for \mbox{$f(R)$ gravity} have been calculated recently \cite{wefR} in the extended de Donder gauge 
\begin{align}
\tensor{\chi}{^\alpha}[g,h]&=-\tensor{g}{^\alpha^\mu}\tensor{g}{^\beta^\nu}\left(\tensor{\nabla}{_\beta}\tensor{h}{_\mu_\nu}-\frac{1}{2}\tensor{\nabla}{_\mu}\tensor{h}{_\beta_\nu}
+\tensor{\Upsilon}{_\beta}\tensor{h}{_\mu_\nu}
\right)\,.\label{ExtDeDonder}
\end{align}
The additional term is linear in
\begin{align}
\tensor{\Upsilon}{_{\mu}}\coloneqq\partial_{\mu}\ln f_1=\frac{f_2}{f_1}\partial_{\mu}R\,.\label{Gam}
\end{align}
The divergent part of the one-loop effective action for \mbox{$f(R)$ gravity} on an arbitrary background reads \cite{wefR}\footnote{The result \eqref{fRAct} in \cite{wefR} has been obtained for the negative of the action \eqref{fRAct}.
Note, however, that \eqref{Gamma1LoopOffShellFinalEPara} is invariant under the change $f\to-f$.}
\begin{widetext}
\begin{align}
\varGamma_1^{f}\big|^{\mathrm{div}}={}&\frac{1}{32\pi^2\varepsilon}\int{\rm d}^4x\,g^{\nicefrac{1}{2}}\left[
-\frac{71}{60}\mathcal{G}
-\frac{609}{80}\tensor*{R}{_\mu_\nu}R^{\mu\nu}
+\frac{1}{3}\frac{f}{f_2}
-\frac{115}{288}\left(\frac{f}{f_1}\right)^2
-\frac{1}{18}\left(\frac{f_1}{f_2}\right)^2
\right.\nonumber\\
&\left.
-\frac{15}{64}\frac{f}{f_1}\,R
+\frac{3919}{1440}R^2
+\frac{15}{64}R\,\Delta\ln f_1
+E\left(\frac{55}{108}E
-\frac{419}{432}\frac{f}{f_1}
+\frac{2933}{864}R
+\frac{221}{288}\Delta\ln f_1\right)\right.\nonumber\\
&\left.-\tensor{E}{_\mu_\nu}\left(\frac{403}{96}\tensor{E}{^\mu^\nu}
+\frac{2987}{288}\tensor*{R}{^\mu^\nu}\right)\right]\,,\label{Gamma1LoopOffShellFinalEPara}
\end{align}
\end{widetext}
with the Gauss-Bonnet term
\begin{align}
\mathcal{G}\coloneqq{}&\tensor{R}{_{\mu\nu\rho\sigma}}\tensor{R}{^{\mu\nu\rho\sigma}}-4\tensor*{R}{_{\mu\nu}}\tensor*{R}{^{\mu\nu}}+R^2\,.\label{GaussBonnet}
\end{align}

\section{Transition between $f(R)$ theories and scalar-tensor theories in the Einstein frame}\label{Sec:Transformations}
The action \eqref{fRAct} for \mbox{$f(R)$ gravity} admits a scalar-tensor formulation, where the extra scalar degree of freedom, included in the higher derivative structure of \mbox{$f(R)$ gravity}, becomes manifest.
The transformation can be performed in two steps.
First we introduce an auxiliary scalar field $\chi$, perform a Legendre transformation and represent the action for \mbox{$f(R)$ gravity} as a scalar-tensor theory \eqref{ActJF} in the JF formulation for the JF scalar field $\varphi$.
In a second step, we perform the transformation \eqref{EFTrafo} to the EF formulation \eqref{ActEF}.
In this way, all information about the original function $f(R)$ is encoded in the EF potential \eqref{EFPot} and the EF field $\hat{\varphi}$.

Starting from the action \eqref{fRAct}, we introduce the auxiliary scalar field $\chi$ and perform a Legendre transformation
\begin{align}
S_{\mathrm{aux}}[g,\chi]=-\int{\rm d}^4x\,g^{\nicefrac{1}{2}}\left[f(\chi)+f_{1}(\chi)(R-\chi)\right]\,.\label{Saux}
\end{align}
Extremizing \eqref{Saux} with respect to $\chi$ leads to the equation
\begin{align}
f_{2}\left(R-\chi\right)=0\,.
\end{align}
For $f_{2}\neq0$ this implies 
\begin{align} 
\chi=R\,.\label{OnShellChiR}
\end{align}
Therefore, ``on-shell'' the action  \eqref{Saux} is equivalent to the original action \eqref{fRAct}.
We define the scalar function $U(\varphi)$
\begin{align}
U(\varphi)\coloneqq f_{1}(\chi)\,.\label{Uchi}
\end{align}
Given a function $f(\chi)$, this relation has to be inverted and explicitly solved for $\chi(\varphi)=\chi(U(\varphi))$.
In terms of \eqref{Uchi}, the action \eqref{Saux} acquires the form of a scalar-tensor theory \eqref{ActJF} with $G(\varphi)=0$,\footnote{Therefore, \mbox{$f(R)$ gravity} corresponds to a subclass of scalar-tensor theories with non-minimal coupling $U(\varphi)$ to gravity without canonical kinetic term, i.e. $G=0$.}
\begin{align}
S_{\mathrm{aux}}[g,\varphi]=\int{\rm d}^4x\,g^{\nicefrac{1}{2}}\left[-U(\varphi)R+V(\varphi)\right]\,.\label{SJF}
\end{align}
The JF potential is given by
\begin{align}
V(\varphi)\coloneqq U(\varphi)\chi(\varphi)-f(\chi(\varphi))\,.\label{Vchi}
\end{align}
Using \eqref{EFTrafo} with $G=0$, we obtain the EF scalar-tensor formulation \eqref{ActEF} for \mbox{$f(R)$ gravity}. 

In order to compare the different formulations, we provide the explicit transformations that bring the EF scalar-tensor theory back to its corresponding \mbox{$f(R)$ formulation}.
First, we present the transformations for the scalar field and its derivatives as well as for the scalar field potential and its derivatives.
The special property $G=0$ of \mbox{$f(R)$ theories} allows to immediately integrate the differential relation \eqref{EFTrafo}.
Using \eqref{Uchi}, we express the EF field $\hat{\varphi}$ in terms of the scalar curvature $R$, 
\begin{align}
\hat{\varphi}(R)=(3\,U_0)^{\nicefrac{1}{2}}\ln\,f_1(R)\,.\label{phifR}
\end{align}
This implies the relation
\begin{align}
\frac{\partial R}{\partial \hat{\varphi}}=(3\,U_0)^{-\nicefrac{1}{2}}\frac{f_1}{f_2}\,.\label{ChainRule}
\end{align}
Combining \eqref{phifR} with \eqref{Gam}, we obtain
\begin{align}
\partial_{\mu}\hat{\varphi}=(3\,U_0)^{\nicefrac{1}{2}}\,\tensor{\Upsilon}{_\mu}\,.\label{DHatPhi}
\end{align}
Using \eqref{EFPot}, \eqref{OnShellChiR} and \eqref{Uchi}, the EF potential can be expressed as a function of scalar curvature $R$,
\begin{align}
\hat{V}(R)=U_0^2\,\frac{R\,f_1-f}{f_1^2}\,.\label{EFPotf}
\end{align}
With \eqref{ChainRule}, we find for the first and second derivatives
\begin{align}
\hat{V}_{1}(R)={}&
U_0^2\,(3\,U_0)^{-\nicefrac{1}{2}}\left[\frac{2\,f-R\,f_1}{\left(f_1\right)^2}\right],\\
\hat{V}_2(R)={}&\frac{U_0}{3f_2}\left[\frac{\left(f_1\right)^2+R\,f_1\,f_2-4f\,f_2}{\left(f_1\right)^2}\right].\label{VPPTrafo}
\end{align}
Second, we collect the conformal transformation rules.
Combing \eqref{EFTrafo} with \eqref{OnShellChiR} and \eqref{Uchi}, we find
\begin{align}
\tensor{\hat{g}}{_\mu_\nu}={}&\frac{f_1}{U_0}\tensor{g}{_\mu_\nu}\,,\label{ConfTrafoMetric}\\
\tensor{\hat{g}}{^\mu^\nu}={}&\frac{U_0}{f_1}\tensor{g}{^\mu^\nu}\,,\label{ConfInvMetric}\\
\hat{g}^{\nicefrac{1}{2}}={}&\left(\frac{f_1}{U_0}\right)^2g^{\nicefrac{1}{2}}\,,\\
\tensor{\hat{\Gamma}}{^{\lambda}_{\mu\nu}}={}&\tensor{\Gamma}{^{\lambda}_{\mu\nu}}+\tensor*{\delta}{^{\lambda}_{(\mu}}\tensor{\Upsilon}{_{\nu)}}-\frac{1}{2}\tensor{g}{_{\mu\nu}}\tensor{\Upsilon}{^{\lambda}}\,,\label{ConfGamma}\\
\tensor{\hat{R}}{^\lambda_\mu_\nu_\rho}={}&\tensor{R}{^\lambda_\mu_\nu_\rho}
-\frac{1}{2}\tensor*{\delta}{^\lambda_{[\nu}}\tensor{g}{_{\rho]}_{\mu}}\tensor{g}{^\alpha^\beta}\tensor{\Upsilon}{_\alpha}\tensor{\Upsilon}{_\beta}\nonumber\\
&
+\frac{1}{2}\left(\tensor*{\delta}{^\lambda_{[\nu}}\tensor{\Upsilon}{_{\rho]}}\tensor{\Upsilon}{_\mu}
-\tensor{g}{^\lambda^\alpha}\tensor{g}{_\mu_{[\nu}}\tensor{\Upsilon}{_{\rho]}}\tensor{\Upsilon}{_\alpha}\right)\nonumber\\
&-\left(\tensor*{\delta}{^\lambda_{[\nu}}\tensor{\nabla}{_{\rho]}}\tensor{\Upsilon}{_{\mu}}-\tensor{g}{^\lambda^\alpha}\tensor{g}{_\mu_{[\nu}}\tensor{\nabla}{_{\rho]}}\tensor{\Upsilon}{_\alpha}\right)\,,\\
\tensor{\hat{R}}{_\mu_\nu}={}&\tensor*{R}{_\mu_\nu}-\frac{1}{2}\tensor{g}{_\mu_\nu}\tensor{g}{^\alpha^\beta}\left(\tensor{\Upsilon}{_\alpha}\tensor{\Upsilon}{_\beta}+\tensor{\nabla}{_\alpha}\tensor{\Upsilon}{_\beta}\right)\nonumber\\
&+\frac{1}{2}\tensor{\Upsilon}{_\mu}\tensor{\Upsilon}{_\nu}-\tensor{\nabla}{_{\mu}}\tensor{\Upsilon}{_{\nu}}\,,\\	
\hat{R}={}&\frac{U_0}{f_1}\left[R-\frac{3}{2}\tensor{g}{^\alpha^\beta}\left(\tensor{\Upsilon}{_\alpha}\tensor{\Upsilon}{_\beta}+2\tensor{\nabla}{_\alpha}\tensor{\Upsilon}{_\beta}\right)\right]\,,
\end{align}
In particular, combining \eqref{DHatPhi} with \eqref{ConfInvMetric} and \eqref{ConfGamma}, the Laplacian of the EF scalar field transforms as
\begin{align}
\hat{\Delta}\hat{\varphi}
={}&(3\,U_0)^{\nicefrac{1}{2}}\frac{U_0}{f_1^2}\Delta f_1\,.\label{LapHatPhi}
\end{align}

\section{Comparison}\label{Sec:Comparison}
Using the explicit transition formulas, provided in the last section, we transform all quantities in the EF formulation to the corresponding expressions in the \mbox{$f(R)$ formulation} and compare them at the classical and quantum level.
For the explicit transformations \eqref{phifR}~--~\eqref{LapHatPhi} not to be singular, we require $f_1\neq0$.
Moreover, for \eqref{Uchi} to be invertible, we require $f_2\neq0$.\footnote{The trivial case $f_2=0$ corresponds to the Einstein-Hilbert action with a cosmological constant.}

\subsection{Tree-level comparison}
By construction, the action of the scalar-tensor theory \eqref{ActEF} in the EF parametrization is equivalent to the action of \mbox{$f(R)$ gravity} \eqref{fRAct}, which can be easily verified by applying the transformation laws \eqref{phifR}~--~\eqref{LapHatPhi} to \eqref{ActEF}.

Likewise, the Einstein equation is easily seen to be equivalent to the equation of motion $E_{\mu\nu}=0$ for \mbox{$f(R)$ gravity} by applying \eqref{phifR}~--~\eqref{LapHatPhi} to \eqref{EinsteinKleinGordonEq}.
In addition, the Klein-Gordon equation for the scalar field in \eqref{EinsteinKleinGordonEq} transforms into the trace of the on-shell condition $E=0$, which therefore does not encode any new information.\footnote{A similar result regarding the equivalence of the equations of motion has been obtained in \cite{Chakraborty:2016ydo}.}
In particular, the equivalence of the equations of motion for scalar-tensor theories and \mbox{$f(R)$ gravity} implies that the on-shell condition can be imposed in either formulation.

\subsection{One-loop comparison}
We apply the transformation formulas \eqref{phifR}~--~\eqref{LapHatPhi} to the divergent part of the off-shell one-loop effective action $\hat{\varGamma}_{1}^{\text{\tiny EF}}$, calculated in the EF \eqref{OneLoopGammaDivEF}.\footnote{It can be shown that the gauge condition \eqref{DeDonder} is equivalent to the gauge condition \eqref{ExtDeDonder} by applying the transformations \eqref{EFPotf}~--~\eqref{VPPTrafo} to the background field.} In this way, we express $\hat{\varGamma}_{1}^{\text{\tiny EF}}$ in terms of its \mbox{$f(R)$ formulation} $\varGamma_{1}^{\text{\tiny EF}}$. Subsequent use of the integration by parts identities, provided in Appendix \ref{AppIBP}, allows to write $\varGamma_{1}^{\text{\tiny EF}}$ in terms of the rescaled extremal $E_{\mu\nu}$,\footnote{We independently checked \eqref{Gamma1OffShellEinsteintofR} with the {\tt Mathematica} computer algebra bundle {\tt xAct} \cite{xAct, Brizuela:2008ra, Nutma:2013zea}.}
\begin{widetext}
\begin{align}
\left.\varGamma_1^{\text{\tiny EF}}\right|^{\mathrm{div}}&=\frac{1}{32\pi^2\varepsilon}\int{\rm d}^4x\,g^{\nicefrac{1}{2}}\left[
-\frac{71}{60}\mathcal{G}
-\frac{609}{80}\tensor*{R}{_\mu_\nu}R^{\mu\nu}
+\frac{1}{3}\frac{f}{f_2}
-\frac{115}{288}\left(\frac{f}{f_1}\right)^2
-\frac{1}{18}\left(\frac{f_1}{f_2}\right)^2
\right.\nonumber\\
&\left.
-\frac{15}{64}\frac{f}{f_1}\,R
+\frac{3919}{1440}R^2
+\frac{15}{64}R\,\Delta\ln f_1+E\left(-\frac{1}{3}E
+\frac{47}{48}\frac{f}{f_1}
+\frac{1}{18}\frac{f_1}{f_2}
+\frac{695}{288}R
+\frac{117}{32}\Delta\ln f_1\right)\right.\nonumber\\
&\left.-\tensor{E}{_\mu_\nu}\left(\frac{331}{96}\tensor{E}{^\mu^\nu}
+\frac{331}{32}\tensor*{R}{^\mu^\nu}\right)\right]\,.\label{Gamma1OffShellEinsteintofR}
\end{align}
\end{widetext}
Note that there is one additional structure in \eqref{Gamma1OffShellEinsteintofR}, proportional to $E\,f_1/f_2$, which is not present in \eqref{Gamma1LoopOffShellFinalEPara}.
Comparing \eqref{Gamma1OffShellEinsteintofR} to the off-shell one-loop divergences \eqref{Gamma1LoopOffShellFinalEPara}, obtained directly for \mbox{$f(R)$ gravity}, we find that the two off-shell results do not coincide. The difference is given by
\begin{align}
&\varGamma_1^{f}\big|^{\mathrm{div}}-\varGamma_1^{\text{\tiny EF}}\big|^{\mathrm{div}}\nonumber\\
={}&\frac{1}{32\pi^2\varepsilon}\int{\rm d}^4x\,g^{\nicefrac{1}{2}} E_{\mu\nu}\left[-\frac{3}{4}\tensor{E}{^\mu^\nu}
-\frac{1}{36}\tensor*{R}{^\mu^\nu}+\left(\frac{91}{108}E\right.\right.\nonumber\\
&\left.\left.
+\frac{53}{54}R-\frac{421}{216}\frac{f}{f_1}-\frac{1}{18}\frac{f_1}{f_2}
-\frac{26}{9}\Delta\ln f_1\right)g^{\mu\nu}\right].
\label{OffShellGamOneLoopComp}
\end{align}
Independent of the choice for the scalar function $f$, the difference between the off-shell divergences never vanishes due to terms proportional to $\tensor{R}{_{\mu\nu}}\tensor{R}{^{\mu\nu}}$ in the first line of \eqref{OffShellGamOneLoopComp}. 
It is clear that the non-equivalence is a pure off-shell effect, as the difference \eqref{OffShellGamOneLoopComp} vanishes on-shell $E_{\mu\nu}=0$. 
Therefore, on-shell, the one-loop divergences for \mbox{$f(R)$ gravity} and its scalar-tensor formulation in the EF are equivalent at the quantum level.

\section{Conclusion}\label{Sec:Conclusion}
We have investigated the quantum equivalence of \mbox{$f(R)$ theories} and scalar-tensor theories by explicitly comparing the one-loop divergences in both formulations for arbitrary background fields.
We find that the off-shell one-loop divergences are ambiguous, as they depend on the formulation, while their on-shell reduction is not.
Our on-shell agreement also provides a strong independent check of the on-shell structures in the result for the one-loop divergences of \mbox{$f(R)$ gravity} obtained in \cite{wefR}.  

On-shell equivalence of $f(R)$ gravity and scalar-tensor theories has also been found in \cite{Bamba:2014mua} for certain cosmological models on a de Sitter background.
The equivalence of \mbox{$f(R)$ gravity} and Brans-Dicke theory has been studied previously in the context of the exact renormalization group \cite{Benedetti:2013nya}.
Although we do not fully agree with their interpretation of the result, their conclusion also seems to support the statement that the off-shell divergences depend on the formulation. 
A similar result has been obtained in \cite{Kamenshchik:2014waa}, where the quantum equivalence of scalar-tensor theories in the JF and EF formulation has been analyzed. 
There, it has been found that the off-shell divergences are parametrization dependent while on-shell the equivalence is retained. 
This on-shell equivalence is to be expected on the grounds of formal equivalence theorems \cite{Chisholm:1961tha,Kamefuchi:1961sb,Coleman:1969sm,Kallosh:1972ap,Tyutin:1982ed}.

The off-shell non-equivalence is not a physical effect but a defect of the underlying mathematical formalism.
The significance of this problem in cosmology might be best explained in the context of inflationary models.
In general, off-shell UV divergences lead to running couplings, whose Renormalization Group (RG) flow is controlled by the corresponding beta functions.
Any ambiguity in the off-shell divergences will therefore induce a corresponding ambiguity in the beta functions and consequently an ambiguity in the values of the coupling constants.
The ambiguity of the beta functions and the couplings is not yet a real problem because neither of them are physical observables. 
The problem in inflationary cosmology arises when these running couplings are evaluated at the energy scale of inflation and simply inserted into the cosmological parameters. These parameters, which inherit the ambiguity from the coupling constants, are then compared to observational data.\footnote{Such a procedure has been applied e.g. in the RG improvement of non-minimal Higgs-inflation \cite{Bezrukov:2008ej,DeSimone:2008ei,Barvinsky:2009fy,Bezrukov:2009db}, which turned out to be crucial for the numerical predictions. In \cite{Steinwachs:2013tr,Kamenshchik:2014waa} it was therefore proposed to use Vilkovisky's unique effective action, see e.g. \cite{Moss:2014nya,Bounakis:2017fkv} for the application of this idea in the context of non-minimal Higgs inflation.}
But what meaning does such a procedure really have?

In order to obtain reliable predictions, it seems to be of crucial importance to define unambiguous cosmological quantum observables, which are in particular manifestly gauge and parametrization independent.

\begin{acknowledgments}
M. S. R. acknowledges financial support from the Deutschlandstipendium. 
\end{acknowledgments}


\appendix

\section{Integration by parts identities}\label{AppIBP}
We can express the $\tensor{\Upsilon}{_{\mu}}$-dependent invariants in terms of $\tensor{E}{_\mu_\nu}$ and its trace $E$ by the following set of identities derived in \cite{wefR},
\begin{align}
\tensor{\Upsilon}{^{\mu;\nu}}={}&\tensor{E}{^\mu^\nu}-\frac{1}{3}\left(E+R-\frac{1}{2}\frac{f}{f_1}\right)\tensor{g}{^\mu^\nu}\nonumber\\
&+\tensor*{R}{^\mu^\nu}-\tensor{\Upsilon}{^\mu}\tensor{\Upsilon}{^\nu}\,,\label{CDGamextremal}
\end{align}
\begin{align}
(\tensor{\Upsilon}{_\mu}\tensor{\Upsilon}{^\mu})^2\overset{\bullet}{=}{}&-\frac{1}{3}\left(E+R-\frac{f}{f_1}\right)(\tensor{\Upsilon}{_\mu}\tensor{\Upsilon}{^\mu})\nonumber\\
&+\frac{2}{3}\left(\tensor{E}{_\mu_\nu}+\tensor*{R}{_\mu_\nu}\right)\tensor{\Upsilon}{^\mu}\tensor{\Upsilon}{^\nu}\,,\\\nonumber\\
\frac{f_1}{f_2}(\tensor{\Upsilon}{_\mu}\tensor{\Upsilon}{^\mu})\overset{\bullet}{=}{}&R\,\Delta\ln f_1\,,\\\nonumber\\
\frac{f}{f_1}(\tensor{\Upsilon}{_\mu}\tensor{\Upsilon}{^\mu})
\overset{\bullet}{=}{}&-\frac{1}{6}\left(E+R-2\frac{f}{f_1}\right)\frac{f}{f_1}\nonumber\\\nonumber\\
&+\frac{1}{2}R\,\Delta\ln f_1\,,\\\nonumber\\
R\,(\tensor{\Upsilon}{_\mu}\tensor{\Upsilon}{^\mu})
\overset{\bullet}{=}{}&-\frac{1}{3}\left(E+R-2\frac{f}{f_1}\right)R\nonumber\\
&+R\,\Delta\ln f_1 \,,\\\nonumber\\
E\,(\tensor{\Upsilon}{_\mu}\tensor{\Upsilon}{^\mu})\overset{\bullet}{=}{}&-\frac{1}{3}\left(E+R-2\frac{f}{f_1}\right)E\nonumber\\
&+E\,\Delta\ln f_1 \,,\\\nonumber\\
\tensor{E}{_\mu_\nu}\tensor{\Upsilon}{^\mu}\tensor{\Upsilon}{^\nu}
\overset{\bullet}{=}{}&\frac{1}{2}\tensor{E}{_\mu_\nu}\tensor{E}{^\mu^\nu}
+\frac{1}{2}\tensor{E}{_\mu_\nu}\tensor*{R}{^\mu^\nu}\nonumber\\
&-\frac{1}{6}E\left(E+R-\frac{1}{2}\frac{f}{f_1}\right)\,,\\\nonumber\\
\tensor*{R}{_\mu_\nu}\tensor{\Upsilon}{^\mu}\tensor{\Upsilon}{^\nu}\overset{\bullet}{=}{}&\tensor{E}{_\mu_\nu}\tensor*{R}{^\mu^\nu}+\tensor*{R}{_\mu_\nu}\tensor*{R}{^\mu^\nu}\nonumber\\
&-\frac{R}{3}\left(E+R-\frac{1}{2}\frac{f}{f_1}\right)\nonumber\\
&+\frac{1}{2}R\,\Delta\ln f_1\,.
\end{align}


\bibliography{STfRV2}
\end{document}